# Synthetic RAW data generator for ESA HARMONY mission

Goulven Monnier
Scalian DS
2 rue Antoine Becquerel,
35700 Rennes, France
goulven.monnier@scalian.com

Benjamin Camus
Scalian DS
2 rue Antoine Becquerel
35700 Rennes, France
benjamin.camus@scalian.com

Yann-Hervé Hellouvry
Scalian DS
2 rue Antoine Becquerel
35700 Rennes, France
yann-herve.hellouvry@scalian.com

Pierre Dubois
Collecte Localisation Satellite
Ramonville-St-Agne, France
pdubois@groupcls.com

Erik De Witte
European Space Agency (ESA)
ESTEC, the Netherlands
erik.de.witte@esa.int

*Abstract*—**In this paper, we introduce HEEPS/MARE, the end-to-end simulator developed for the SAR oceanographic products of ESA Earth Explorer 10 mission, Harmony, expected to launch in Decembre 2029. Harmony is primarily dedicated to the observation of small-scale motion and deformation fields of the Earth surface (oceans, glaciers and ice sheets, solid Earth), thanks to passive SAR/ATI receivers carried by two companion satellites for Sentinel-1. The paper focuses on the raw data generator designed to efficiently simulate large, heterogeneous, moving oceanic areas and produce the acquired SAR/ATI bistatic IQ signals. The heterogeneous sea-surface model, bistatic scattering model, multi-GPU implementation and achieved performance are emphasized. Finally, sample results are presented, to illustrate the ability of Harmony to map wind and surface current vectors at kilometric scale.**

## I. INTRODUCTION

Expected to launch in December 2029, Harmony is a constellation of two SAR (Synthetic Aperture Radar) equipped satellites operated by the European Space Agency (ESA) as an Earth Explorer mission, which will run in tandem with a Sentinel-1 satellite. They will monitor changes in the Earth's surface, as well as monitor ocean surface conditions such as wind, currents, and temperature. Each Harmony satellite will carry two instruments on board: a receive-only C-band SAR (Synthetic Aperture Radar) that will detect C-band radio waves emitted from the SAR of a Sentinel-1 satellite, and a multibeam thermal-infrared instrument.

Harmony satellites will fly alternatively in two formations dedicated to Across-track- and Along-Track interferometry (ATI and XTI).

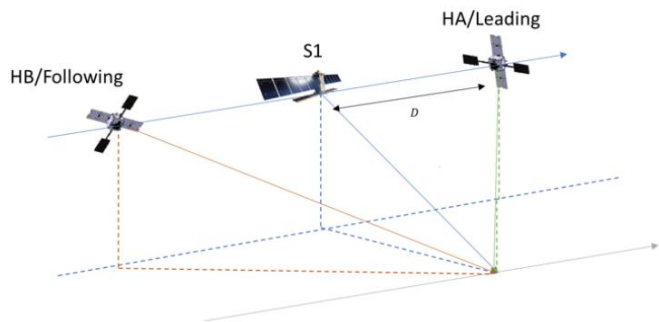

*Figure 1: Harmony A (HA) and Harmony B (HB) satellites in ATI formation, preceding and following Sentinel-1 (S1) at distance D~200-300 km. All 3 SAR antennas mechanically point to the same surface target.*

This paper is about HEEPS/MARE, the E2E simulator developed for performance assessment of the SAR ocean products derived from data acquired in ATI formation (see Figure 1). In this configuration, each Harmony performs ATI thanks to its antenna divided into up to three phase centers, thus giving access to radial velocities including contributions from surface waves and current. Wave spectra are retrieved from imaged NRCS and Doppler modulations, allowing the wave Doppler to be estimated and subtracted, leaving only the surface current contribution. Combining the two Harmonys LOS finally gives access to the Total Surface Current Vector (TSCV) with an expected error an error lower than 0.1 m/s at a resolution of a few km, which constitutes the mission main L2 product.

## II. HEEPS/MARE END-TO-END SIMULATOR

### A. Overview

The ocean products performance, derived in the early development phase with models and targeted simulations, was consolidated through the use of end-to-end (E2E) simulations, following ESA's proven process for Earth Explorer selection. The E2E simulators are classic tools for characterizing the performance of a mission, as defined by the science requirements. They integrate the definition of a set of geophysical truths, the geometry and timing of the acquisition, the transfer function of the instrument, and the prototyping of all levels of processing (On-board L0, L1, L2). At the end of the L2 processing, the estimates of the geophysical parameters of interest can be compared to the geophysical truth sets used as an input to the simulation.

In the frame of Harmony phase A, the development and operation of the HEEPS/MARE end-to-end simulator was entrusted to a consortium formed by CLS (in charge of the Geometry Module and L1c-L2 processing) , Scalian (in charge of the Scene-and-Instrument Module, and the L0 raw data generation) and NORCE (in charge of L1a and L1b processing). CLS and SCALIAN were able to draw on similar experience in the frame of SKIM mission (EE9 candidate). Contrary to SKIM, Harmony involves SAR and ATI processing (in a squinted, bistatic context), hence the need for expertise in this field provided by NORCE. In the following, this paper concentrates on the Scene-and-Instrument Module (SIM) developed by SCALIAN.

*B. Scene and Instrument module*

SIM is a raw data generator, taking as inputs descriptors of the instruments, the geometry, the acquisition mode and the geophysical scene. Although the scene and the instrument are usually supported by two distinct modules in E2E simulators, they have been merged in HEEPS, mainly because the scene is dynamically generated where and when it is actually sensed by the instruments. Only this allows handling efficiently the huge number of scatterers needed for representing accurately the sea surface topography and kinematics. More generally, the SIM module design was mostly guided by the search for computational performance.

The need for performance results from the chosen modeling approach, consisting in imitating, as far as possible, the measuring process. Hence, the sea surface generated in the footprint corresponding to each receiving channel is composed of an assembly of independent moving scatterers (facets). The received signals are computed as sums of scatterers complex amplitudes, weighted by emitter and receiver antenna patterns. The facets size is chosen as a tradeoff between targeted resolution of SAR images, assumptions of the EM scattering asymptotic model and computing burden. The size of the facets also governs the fraction of the surface Doppler being supported by their motion. The remaining fraction, related to unresolved intra-facet kinematics, has to be modelled and added separately. A correct representation of the surface Doppler spectrum is crucial, as it governs the accessible ATI performance through the coherence time (Frasier et al., 2001).

This straightforward modelling scheme allows accounting not only for realistic instrument nominal characteristics (orbit, attitude, antenna patterns…) but also for their known or unknown deviations from nominal values, as expected from a performance assessment tool. Moreover, it provides synthetic raw data which can be ingested by processors exactly as real ones would be. This is at the cost of a heavy computational burden: individual simulations are expensive and multiple simulations are required given their deterministic nature. Therefore, ensuring the simulator to be practically useful requires High Performance Computing.

### III. SEA SURFACE AND EM SCATTERING MODEL

The EM wave scattering by the ocean surface is modelled through generating a realization of the moving sea surface at metric resolution and accounting for unresolved intra-facet roughness through statistical modelling. This approach resorts to a "two-scale" model but should be distinguished from the two-scale method initially developed by Valenzuela [1], who considered Bragg waves modulated by long gravity waves. In this perspective, we call it "Two scale Grid Model", referring to the gridded surface whose resolution determines the practical partitioning between long and short waves.

The sea surface is divided into square tiles, which are assigned a wave spectrum whose low frequency side, describes the resolved waves. The tile topography is used generated through the *Choppy Wave Model* (CWM) [2]. The CWM produces a skewed distribution of the surface height and, as a quasi-lagrangian model, makes the grid points follow waves orbital motion, which is appropriate for Doppler simulations. However, if a linear surface is preferred, the user is free to deactivate the CWM option.

To allow generating large non-periodic or heterogeneous surfaces sea states (i.e. spatially varying wave frequency spectrum), the tiles are not used as-is, but combined linearly. The surface is dived into heterogeneous square tiles TL of size L, in which four homogeneous tiles $T2L_i$ (i=1…4) of size 2L overlap. The surface displacement at a grid point in TL is computed as a weighted sum of the displacements from the four overlapping homogeneous T2L tiles (Figure 2):

$$\boldsymbol{D}_{TL}(\bar{\mathbf{x}}, t) = \sum_{i=1}^{4} W^i(\bar{\mathbf{x}}) \boldsymbol{D}_{T2L}^i(\bar{\mathbf{x}}, t) \quad (1)$$

Where $\boldsymbol{D}_{TL}(\bar{\mathbf{x}}, t)$ is the displacement computed on the TL tile at position at rest $\bar{\mathbf{x}}$ and time t, $\boldsymbol{D}_{T2L}^i(\bar{\mathbf{x}}, t)$ are the displacements of the four overlapping T2L tiles at same location end time. $W^i(\bar{\mathbf{x}})$ are constant weights. They are chosen so that at TL corner I, the frequency spectrum corresponds exactly to $S_i$, the wavenumber spectrum used for generating the homogeneous tile $T2L_i$.

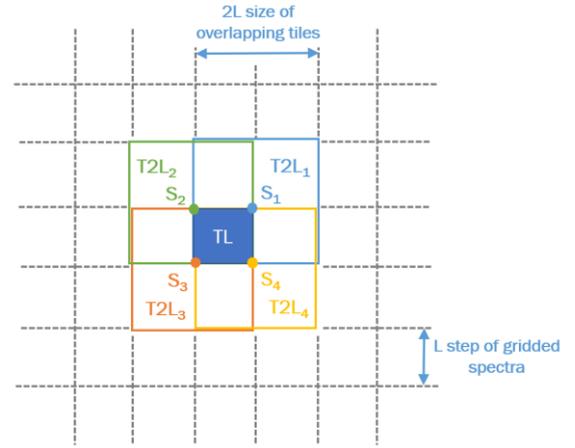

*Figure 2: generating one heterogeneous TL tile from 4 homogeneous T2L tiles.*

The spectra $S_i$ are thus specified on a grid of step L, which can be set by the user but is typically on the order of a few kilometers, which defines the finest scale at which the sea state can vary. In case of a uniform sea state, the spectrum is constant, but the phases and amplitudes are still randomly generated at each T2L tile, with no correlation between tiles. It can then be seen that combining the tiles displacements forces the surface autocorrelation, which gets location-dependent, to fall to zero at distances varying from L to 2L. This is not a problem if the wave correlation length, related to the spectrum width, is much shorter than L, which is ensured provided the swell spectrum is well resolved.

As the weighted sum is carried on uncorrelated displacements and must preserve, on average, the total wave energy, the squares of the weighting coefficients must add to unity. This condition, together with having one unit weight at each TL corner, are easily fulfilled by choosing:

$$W^i(x, y) = \left(1 - w_x^i - w_y^i + w_x^i \cdot w_y^i\right)^{1/2}$$
$$w_x^i = \frac{|x^i - x|}{L}, \quad w_y^i = \frac{|y^i - y|}{L} \quad (2)$$

where $\bar{x} = (x, y)$ are the coordinates in the heterogeneous tile and $(x^i, y^i)$ are the coordinates of corner $i$, center of tile T2L$_i$

This simple scheme allows generating strongly heterogeneous wave fields, like that presented in Figure 3 where the direction, wavelength and azimuthal spread of the swell vary sharply from the left to the right. However, it should be noted that no physical consistency is ensured, as the scheme only resorts to spectrum interpolation. Hence the need for an external description of the sea state, which is translated into surface realization.

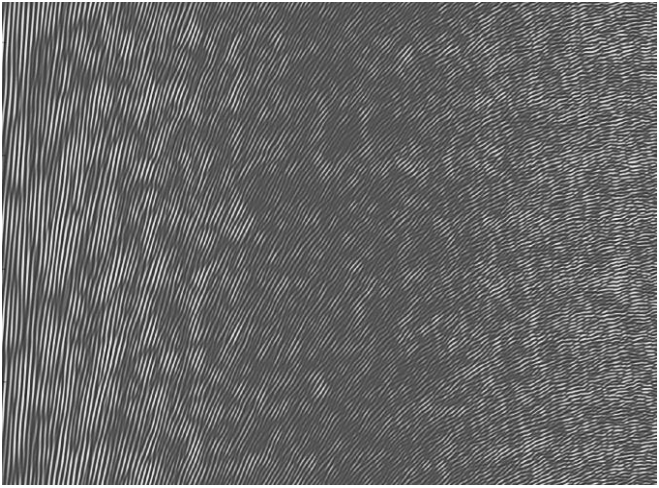

*Figure 3: an example of heterogeneous swell field generated through the algorithm described above.*

The low-frequency part of the wave spectrum corresponding to resolved waves determines the surface topography, hence the orientation of the facets. The scattering matrix is then computed at the facet level, given its orientation and its spectral content, given by the high-frequency side of the surface spectrum.

*A. Bistatic EM scattering model*

In the TSGM, the received instantaneous signal is essentially obtained by summing the complex contributions from all the facets lying in a given radar pixel. The field contribution at the receiver from the facet identified reads:

$$E_t^q(t_r) \propto \frac{E_t^p(t_t)}{r_t(t_t) r_r(t_r)} \sqrt{A} S^{pq}(\mathbf{k_s}, \mathbf{k_i}, t) e^{-iK_0(r_t(t_t) + r_r(t_r) + v_c t)} \quad (3)$$

Where $E_t^p(t_t)$ and $E_t^q(t_r)$ are the transmitted and received fields according linear polarizations $p$ and $q$, at transmission and reception times $t_t$ and $t_r$. $r_t(t_t)$ and $r_r(t_r)$ are the corresponding ranges from the antenna phase center to the considered facet. Also included in the contribution to the phase of the surface current of velocity $\mathbf{V_c}$:

$$v_c = \mathbf{V_c} \cdot (\hat{r}_t + \hat{r}_r) \quad (4)$$

The facet scatterer is described through its area $A$ and normalized complex reflection coefficient $S^{pq}(\mathbf{k_s}, \mathbf{k_i}, t)$ where $\mathbf{k_i}$ and $\mathbf{k_s}$ refer to the wave vectors of incident and scattered waves. In this expression appears the geometrical phase term $K_0(r_t(t_t) + r_r(t_r))$ which accounts for the resolved surface topography. The phase from the unresolved waves is thus accounted for through $S^{pq}(\mathbf{k_s}, \mathbf{k_i}, t)$, where the time-dependency takes in charge the finite correlation time of the facet contribution (i.e. its Doppler width).

In the case of Harmony, capturing the polarization effects related to the bistatic geometry is of primary importance. The simplest scattering model offering this capability with acceptable validity in the moderate incidence domain is SSA1 [3]. It also allows a very efficient implementation. SSA2 would arguably be more accurate [4], particularly at catching the polarization effects related to the waves, but it has been left apart for now, given its much more intricate implementation.

In the frame of a first-order scattering model like SSA1, the time-dependent complex reflection coefficient may be written:

$$S^{pq}(\mathbf{k_s}, \mathbf{k_i}, t) = B_1^{pq}(\mathbf{k_s}, \mathbf{k_i}) \sqrt{KI_s(\mathbf{Q})} \, s(t) \quad (5)$$

where $B_1^{pq}(\mathbf{k_s}, \mathbf{k_i})$ is the first order Bragg scattering kernel, $KI_s(\mathbf{Q})$ and is the Kirchhoff integral over short waves given by:

$$KI_s(\mathbf{Q}) = \frac{1}{\pi Q_z^2} \int \left( e^{-Q_z^2(\rho_s(0) - \rho_s(x))} - e^{-Q_z^2 \rho_s(0)} \right) e^{-i\mathbf{Q_H} \cdot \mathbf{x}} d\mathbf{x} \quad (6)$$

$\mathbf{Q} = \mathbf{k_s} - \mathbf{k_i}$ is the Ewald vector with horizontal and vertical components $\mathbf{Q_H}$ and $Q_z$ in the facet frame and $\rho_s(x)$ is the short waves bidimensional height autocorrelation function, which is the inverse Fourier transform of the intra-facet side of the wave spectrum.

$s(t)$, the time-dependent complex intra facet speckle reads:

$$s(t) = [R(t) + iI(t)] e^{2i\pi f_s t} \quad (7)$$

Where the intra-facet Doppler shift $f_s$ can be related to $KI_s$ by:

$$f_s = -\frac{i}{2\pi} \frac{\partial_t KI_s}{KI_s} \quad (8)$$

$R(t = 0)$ and $I(t = 0)$ are independent normally distributed random numbers with zero mean and variance $1/2$, ensuring that $\langle s(0) s^*(0) \rangle = \langle |R(0)|^2 \rangle + \langle |I(0)|^2 \rangle = 1$. Ideally, their time dependance should ensure that the time correlation function is of the form:

$$\langle s(0) s^*(t) \rangle = e^{-\frac{t^2}{2\tau_s^2}} e^{-2i\pi f_s t} \quad (9)$$

Where the correlation time $\tau_c$ is related to the intra-facet Doppler width $\sigma_s$ through $\tau_s = (2\pi\sigma_s)^{-1}$. Achieving such a gaussian time correlation function at the facet level is computationally prohibitive. A practical alternative is thus adopted, consisting of using an order-1 autoregressive model, leading to an exponential form $e^{-\frac{t}{\tau_s}}$:

$$R(t + \Delta t) = \beta R(t) + (1 - \beta^2)^{\frac{1}{2}} R' \\ \beta = e^{-\frac{\Delta t}{\tau_s}}, \quad R' \sim N(\mu = 0, \sigma = 1/2) \quad (10)$$

In most cases, the sea surface backscatter correlation time is dominated by the contribution of the resolved waves, acting through the geometrical phase term. Therefore, the precise shape of the intra-facet correlation function is not essential, and it may even be set to unity.

In the formulation described above, the TSGM can be very accurately and efficiently implemented, as the sea-state dependent, computationally intensive calculations of $KI_s$ and $f_D$ can be performed offline to generated look up tables. The polarized, highly geometry-sensitive kernel $B_1^{pq}(\mathbf{k_s}, \mathbf{k_i})$ is computed exactly online. Computing the scattered field components also involves two changes of polarization basis, first from the emitter to the facet, then to the receiver frame.

A convenient way of computing, storing and using $KI_s$ and $f_s$ is to resort to azimuthal harmonic expansions of $KI_s$ and $\partial_t KI_s$ of the form:

$$KI_s \simeq a_0 + 2 \sum_{m=1}^{m_{max}} a_{2m} \cos(2m\chi)$$
$$\pi Q_z^2 \, \partial_t KI_s = b_0 + i \sum_{n=0}^{n_{max}} b_{2n+1} \cos((2n+1)\chi) \quad (11)$$

Where the bistatic angle $\chi$ reads (e.g. [5]):

$$\tan \chi = \frac{\mathbf{Q_H} \cdot \hat{\mathbf{y}}}{\mathbf{Q_H} \cdot \hat{\mathbf{x}}} = \frac{\sin \phi_s \sin \theta_s - \sin \phi_i \sin \theta_i}{\cos \phi_s \sin \theta_s - \cos \phi_i \sin \theta_i} \quad (12)$$

The intra-facet Doppler shift is then approximately by:

$$f_s = -\frac{i}{2\pi} \frac{\partial_t KI_s}{KI_s} \simeq -\frac{\sum_{n=0}^{n_{max}} b_{2n+1} \cos((2n+1)\chi)}{2\pi \sum_{m=1}^{m_{max}} a_{2m} \cos(2m\chi)} \quad (13)$$

Coefficients $a_{2m}$ and $b_{2n+1}$ are defined as the following 1D radial integrals:

$$a_0 = 2\pi \int_0^\infty \left[ I_0(Q_z^2 \rho_2) J_0(Q_H r) e^{-Q_z^2(\rho_0(0) - \rho_0(r))} - J_0(Q_H r) e^{-Q_z^2 \rho_0(0)} \right] r dr$$

$$a_{2m} = 2\pi \int_0^\infty I_m(Q_z^2 \rho_2) J_{2m}(Q_H r) e^{-Q_z^2(\rho_0(0) - \rho_0(r))} r dr \quad (14)$$

$$b_{2n+1} = 2\pi Q_z^2 \int_0^\infty [\rho_1(r)(I_{n+1}(Q_z^2 \rho_2(r)) - I_n(Q_z^2 \rho_2(r))) - \rho_3(r)(I_{n+2}(Q_z^2 \rho_2(r)) - I_{n-1}(Q_z^2 \rho_2(r)))] J_{2n+1}(Q_H r) r dr$$

Where $\rho_i(r)$ relates to space and time correlation function and thus to short waves spectrum through the following 1D integrals over $k$:

$$\rho_0(r) = \int_0^\infty \Psi_s(k) J_0(rk) dk$$

$$\rho_1(r) = \int_0^\infty \omega(k) \Psi_s(k) J_1(rk) \left( \frac{4}{\pi} + \frac{4}{3\pi} \Delta_s(k) \right) dk \quad (15)$$

$$\rho_2(r) = \int_0^\infty \Psi_s(k) \Delta_s(k) J_2(rk) dk$$

$$\rho_3(r) = \int_0^\infty \omega(k) \Psi_s(k) J_3(rk) \left( \frac{4}{3\pi} - \frac{12}{5\pi} \Delta_s(k) \right) dk$$

Where $\Psi_s(k)$ and $\Delta_s(k)$ are the omnidirectional short waves spectrum the spreading function as described in [6], and $\omega(k)$ is the dispersion relationship for linear waves. Expressions for $\rho_{2n+1}$ can be found in [7], but the ones obtained here, though slightly different, compare favorably with the reference model.

In HEEPS, coefficients $a_{2m}$ and $b_{2n+1}$ are computed offline through equations (14) and (15), then tabulated as a function of U10, $\|\mathbf{Q}\|/K_0$, $Q_z/K_0$. These two last parameters have been found to allow the most compact lookup tables. During the simulation, the scattering geometry ($\mathbf{Q}$, $\chi$) is computed at the facet level, from which $B_1^{pq}(\mathbf{k_s}, \mathbf{k_i})$ is obtained. Coefficients $a_{2m}$ and $b_{2n+1}$ are then interpolated in the LUT, allowing, together with $\chi$, estimating $KI_s$ (eq. (11)), $f_s$ (eq. (13)), $s(t)$ (eq. (7)) and $S^{pq}(\mathbf{k_s}, \mathbf{k_i}, t)$ (eq. (5)). Performing a rotation of the polarization bases for the incident and scattered waves finally allows computing the scattered field at the receiver (eq. (3)).

*B. Radar equation and instrument model*

Beyond equation (3), which is simplified to focus on the scattering model, the received filed contribution from a facet at location $(x, y)$ may be formally written:

$$E_j^r(x, y) = \frac{e^{-iK_0(r_t(t_t) + r_r(t_r) + v_c t)}}{4K_0 \pi^{1/2} r^t(x, y, t_t) r^r(x, y, t_r)} \sum_i \sqrt{P_i^r} \, G_i^r(x, y) \mathcal{R}_{i,j}(x, y) G_i^t(x, y) \quad (16)$$

With:

- $P_i^r$      power transmitted in channel $i$
- $E_j^t$      Received complex amplitude in channel $j$ (platform, phase centre, polarization)
- $G_i^t, G_j^r$      Tx and Rx complex antenna gains in channels $i$ and $j$, also including atmospheric losses and delays.
- $\mathcal{R}_{i,j}$      Complex linear operator embedding the scattering matrix, polarization basis rotations, facet random amplitude, phase and Doppler (intra-facet waves)

Eq. (16) stresses the role of antenna patterns, which are given for each Tx and Rx phase center and polarization. They must be estimated at each facet and for each pulse, constituting a significant part of the computational burden in the case of Harmony, especially in TOPSAR modes, in which they are time-dependent. Whatever the format they are provided in, antenna patterns are converted into u-v grids at the appropriate resolution and loaded in the GPU memory. Up to know, they where specified by the industrial consortia in the form of antenna phased array models.

The complex amplitudes $E_j^r(x, y)$ of all facets lying in range slices, i.e. verifying

$$R_i \leq r^t(x, y) + r^r(x, y) \leq R_{i+1} \quad (17)$$

are subsequently added to obtain one time sample of the received signal. The range interval $R_{i+1} - R_i$ is small compared to the chirp length, so that the obtained profile can be regarded as the impulse response of the surface and subsequently convolved by the waveform. The generated raw data are output together with the waveforms, so that range compression can be performed by the L1a processor. The thermal noise is provided separately.

IV. IMPLEMENTATION DETAIL

*A. Computation scheme and module architecture*

The SIM module general architecture, given in Figure 4, is mainly determined by the computation scheme specially designed for a multi-GPU implementation. The Internal Preprocessor is in charge of partitioning the simulation into elementary square TL tiles whose contributions are subsequently computed separately on the available GPUs by the Raw Data Generator. They are finally combined by the Postprocessor. The outputs consist of the obtained raw data plus a set of "ground truth" data (extract of surface topography, NRCS, velocity…) provided for validation purpose.

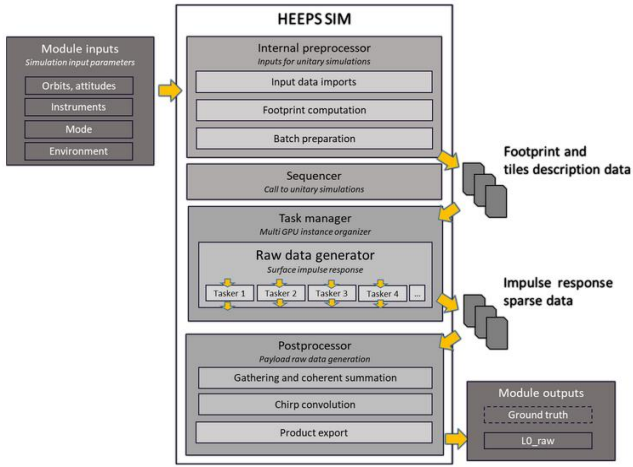

*Figure 4: general organization and interfaces of the Scene and Instrument Module*

The Internal Preprocessor maps a regular grid of tiles on the earth ellipsoid, in a slanted lat-long frame so that the projection deformation remains negligible in the swath. From the geometry and instruments inputs, it determines the tiles which will contribute to the raw data (Figure 5, left, in red) and the list of the corresponding Sentinel 1 transmitted pulses. For a given pulse, only a subset of the selective tiles contribute (Figure 5, left, in green). On Figure 5 (on the right) are shown some of the contributing tiles for pulse 0, each of size 1024X1024 with 2.5 m grid step.

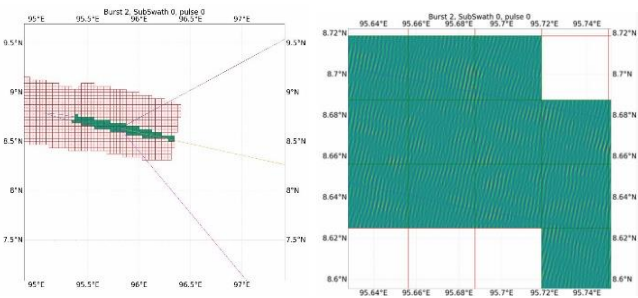

*Figure 5: views from the "ground truth" outputs, showing (left) tiles for one burst (in red) and those contributing for one pulse (in green). On the right: zoom on the contributing tiles showing the swell.*

### B. GPU optimization and performance

The simulation requires a large number of resources and is computationally consuming. To optimize and make the chain scalable with very large scenes, a parallelisation approach is adopted relying on a general-purpose processing on graphics processing units (GPGPU). Due to the limited memory available on these accelerators, the scene (i.e. the facets information) is partitioned into multiple batches corresponding to the TL tiles, which are computed sequentially on the GPU. Then, for each tile and radar pulses, the corresponding surface is generated and the radar equation is applied to compute the scatterers impulse response.

If available on the computing node, the computation can also be distributed on several devices. In this multi-GPU strategy, the available devices divide up the batch of tiles to process. The different tiles are then simulated in parallel on different accelerators. Therefore, each GPU computes only a subpart of the final raw data. Once all the chunks have been processed, a final thread gathers the sparse raw data. This implementation is illustrated in Figure 6.

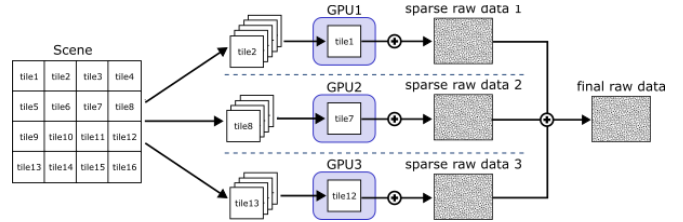

*Figure 6. Illustration of the multi-GPU implementation.*

The following computing times are achieved for 7 receiving channels (1 phase center for Sentinel 1, 3 for both Harmony 1 and 2) on 4 NVIDIA 1080 GTX GPUs:

| S1 Mode | Image size (km) | Dwell time (s) | Computing time (min) |
|---|---|---|---|
| Wave mode | 20x20 | 0.64 | 50 |
| Interferometric Wide Swath (3 subswaths, 1 burst) | 250x20 | 0.14 | 120 |
| Extra Wide Swath (5 subswaths, 1 burst) | 400x20 | 0.07 | 180 |

Note that the computing time is not simply proportional to the illuminated surface but is also determined by the dwell time (much shorter in the TOPSAR modes than in Wave mode) and specific computing costs like that of allocating new tiles on the GPU memory.

A new fast mode is currently tested on nadir altimetry simulation cases, which may further reduce the computing time by a factor of 5 to 20 (depending on the considered mode). It consists in estimating the Doppler spectrum of each range gate at a frequency $PRF/n_{fast}$ smaller than the PRF, then generating the missing pulses through 1D IFFT and linear combination of the resulting signals. In essence, it is very similar to the method used to generated heterogeneous surfaces, but in 1D rather than 2D. In nadir altimetry cases, it proved to be applicable with values of $n_{fast}$ as high as 750, with corresponding speedup factors of ~100 (from 20 hours to 12 minutes). In the case of Harmony, $n_{fast}$ will have to remain much smaller, but significant speedup factors are nevertheless expected.

## V. PROCESSING AND SAMPLE RESULTS

The processing chain from L0 data output by the SIM to L2 products was prototyped by CLS and NORCE and is not in the scope of this paper. The combined use of SAR and ATI in a squinted "doubly bistatic" context makes the successive processing steps very challenging, especially when instrumental errors are considered (e.g. residual ATI baseline). It is worth underlying that the inevitable limitations of the previously described forward model (concerning the ocean surface and the scattering model) were handled in the processing by constructing adapted Geophysical Model Functions (GMFs). The model, although imperfectly reproducing the real world, is thus considered as the truth, so that the final mission performance assessment is not artificially underestimated. In this perspective, an effort was made to model the TGSM so as to derive corresponding GMFs analytically.

We simply present here an overview of the results obtained during Harmony Phase A, which confirmed the ability of the mission to catch wind vectors and Total Surface Current Vectors (TSCV) at high resolution.

The L1a processor comprises range compression, $\omega - k$ focusing, radiometric calibration and coregistration, and production of ATI interferograms (ref NORCE). The L1b-L1c module computes the NRCS and cross-polarization covariance, the Doppler products (ATI and Doppler centroid) and the modulation co- and cross-spectra. Finally, the L2 processor produces the Total Surface Current Vectors (TSCV), the Surface Stress Vector (SSV) and the wave spectra.

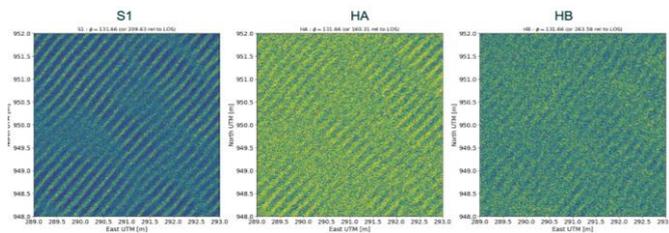

*Figure 7: Focused Wave Mode scene with strong coherent swell: Sentinel 1 (monostatic, left), Harmony A (bistatic, center) and Harmony B (bistatic, right).*

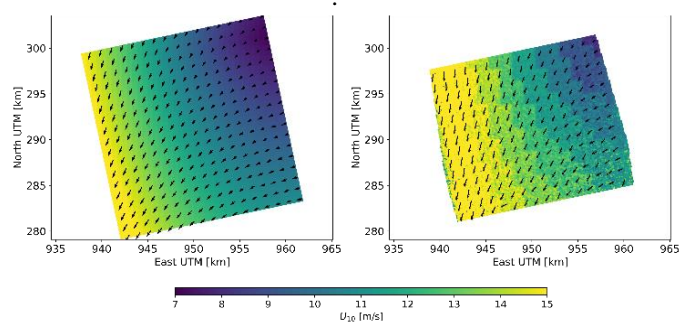

*Figure 8: wind vectors ground truth (on the left) and estimated (on the right) in Wave Mode.*

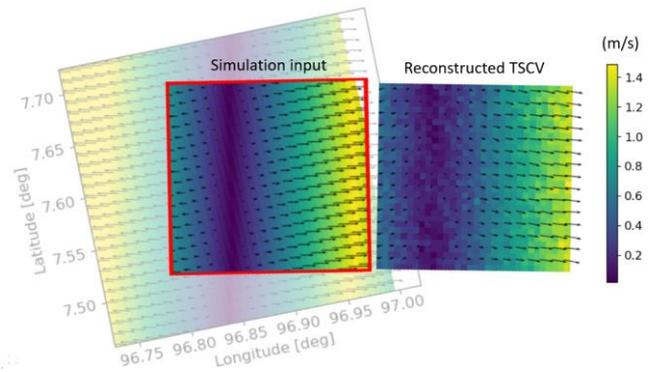

*Figure 9: estimated Total Surface Current Vector (TSCV) in Wave Mode.*

## VI. CONCLUSION

Estimated performance obtained from an early version of HEEPS/MARE were already included in the presentation of Harmony at EE10 UCM, on 5 July 2022 (https://www.esa.int/esatv/Videos/2022/07/Presentations_and_discussions_on_Harmony_EE10_UCM). Synthetic raw data from the SIM module have been used to support the development of a new efficient focusing algorithm for high-squint bistatic SAR data [8]. The E2E simulator will keep on supporting Harmony mission definition phases in the near future, especially through increasingly realistic and accurate instrument and processing models.

### ACKNOWLEDGMENT

HEEPS/MARE was developed under ESA fundings.

### REFERENCES

[1] Valenzuela, G. R. (1968). Scattering of electromagnetic waves from a tilted slightly rough surface. *Radio Science*, *3*(11), 1057-1066.

[2] Nouguier, F., Guérin, C. A., & Chapron, B. (2009). "Choppy wave" model for nonlinear gravity waves. Journal of geophysical research: oceans, 114(C9).

[3] G. Soriano and C.A. Gue´rin, A cutoff invariant two-scale model in electromagnetic scattering from sea surfaces, IEEE Geosci. Remote Sens. Lett. 5 (2008), pp. 199–203.

[4] Fois, F., Hoogeboom, P., Le Chevalier, F., & Stoffelen, A. (2015). An analytical model for the description of the full-polarimetric sea surface D oppler signature. *Journal of Geophysical Research: Oceans*, *120*(2), 988-1015.

[5] Bourlier, C. (2018). *Radar Propagation and Scattering in a Complex Maritime Environment: Modeling and Simulation from MatLab*. Elsevier.

[6] Elfouhaily, Tanos, et al. "A unified directional spectrum for long and short wind-driven waves." *Journal of Geophysical Research: Oceans* 102.C7 (1997): 15781-15796.

[7] Mouche, A. A., Chapron, B., Reul, N., & Collard, F. (2008). Predicted Doppler shifts induced by ocean surface wave displacements using asymptotic electromagnetic wave scattering theories. *Waves in Random and Complex Media*, *18*(1), 185-196.

[8] Yitayew, T. G., Grydeland, T., Larsen, Y., and Engen, G. (2023). Processing of high squint bistatic SAR data: The case of Harmony. SeaSAR